# The Wide Field Infrared Survey Telescope:
## 100 Hubbles for the 2020s

by members of the WFIRST Project and Formulation Science Investigation Teams

Rachel L. Akeson,[1] Lee Armus,[1] Etienne Bachelet,[2] Vanessa P. Bailey,[3] Lisa Bartusek,[4] Andrea Bellini,[5] Dominic J. Benford,[6] David P. Bennett,[4] Aparna Bhattacharya,[7] Ralph C. Bohlin,[5] Martha L. Boyer,[5] Valerio Bozza,[8] Geoffrey Bryden,[3] Sebastiano Calchi Novati,[1] Kenneth G. Carpenter,[4] Stefano Casertano,[5] Ami Choi,[9] David A. Content,[4] Pratika Dayal,[10] Alan M. Dressler,[11] Olivier Doré,[3] S. Michael Fall,[5] Xiaohui Fan,[12] Xiao Fang,[12] Alexei V. Filippenko,[13] Steven L. Finkelstein,[14] Ryan J. Foley,[15] Steven Furlanetto,[16] Jason Kalirai,[17] B. Scott Gaudi,[9] Karoline M. Gilbert,[5] Julien H. Girard,[5] Kevin J. Grady,[4] Jenny E. Greene,[18] Puragra Guhathakurta,[15] Chen Heinrich,[3] Shoubaneh Hemmati,[3] David Hendel,[19] Calen B. Henderson,[1] Thomas Henning,[20] Christopher M. Hirata,[9] Shirley Ho,[21] Eric Huff,[3] Anne Hutter,[10] Rolf A. Jansen,[22] Saurabh W. Jha,[23] Samson A. Johnson,[9] David O. Jones,[14] Jeremy Kasdin,[18] Patrick L. Kelly,[24] Robert P. Kirshner,[25] Anton M. Koekemoer,[5] Jeffrey W. Kruk,[4] Nikole K. Lewis,[26] Bruce Macintosh,[27] Piero Madau,[15] Sangeeta Malhotra,[4] Kaisey S. Mandel,[28] Elena Massara,[21] Daniel Masters,[3] Julie McEnery,[4] Kristen B. W. McQuinn,[23] Peter Melchior,[18] Mark E. Melton,[4] Bertrand Mennesson,[3] Molly S. Peeples,[5] Matthew T. Penny,[9] Saul Perlmutter,[13] Alice Pisani,[18] Andrés A. Plazas,[18] Radek Poleski,[9] Marc Postman,[5] Clément Ranc,[4] Bernard J. Rauscher,[4] Armin Rest,[5] Aki Roberge,[4] Brant Robertson,[15] Steven Rodney,[29] James E. Rhoads,[4] Jason Rhodes,[3] Russell E. Ryan Jr.,[5] Kailash Sahu,[5] Dan Scolnic,[30] David Sand,[12] Anil Seth,[31] Yossi Shvartzvald,[1] Karelle Siellez,[15] Arfon M Smith,[5] David N. Spergel,[21] Keivan Stassun,[32] Rachel A. Street,[2] Louis-Gregory Strolger,[5] Alexander S. Szalay,[33] John Trauger,[3] M. A. Troxel,[30] Margaret Turnbull,[34] Roeland P. van der Marel,[5] Anja von der Linden,[35] Yun Wang,[1] David Weinberg,[9] Benjamin F. Williams,[36] Rogier A. Windhorst,[22] Edward J. Wollack,[4] Hao-Yi Wu,[9] Jennifer C. Yee,[37] Neil T. Zimmerman[4]

[1] IPAC/Caltech
[2] Las Cumbres Observatory
[3] Jet Propulsion Laboratory/Caltech
[4] NASA Goddard Space Flight Center
[5] Space Telescope Science Institute
[6] NASA Headquarters
[7] University of Maryland College Park
[8] Università di Salerno
[9] Ohio State University
[10] Kapteyn Institute, University of Groningen
[11] Carnegie Institution for Science
[12] University of Arizona
[13] U. C. Berkeley
[14] The University of Texas at Austin
[15] U. C. Santa Cruz
[16] U. C. Los Angeles
[17] Applied Physics Laboratory
[18] Princeton University
[19] University of Toronto
[20] Max Planck Institute for Astronomy
[21] Flatiron Institute
[22] Arizona State University
[23] Rutgers University
[24] University of Minnesota
[25] Gordon and Betty Moore Foundation
[26] Cornell University
[27] Stanford University
[28] University of Cambridge
[29] University of South Carolina
[30] Duke University
[31] University of Utah
[32] Vanderbilt University
[33] Johns Hopkins University
[34] SETI Institute
[35] Stony Brook University
[36] University of Washington
[37] Harvard-Smithsonian Center for Astrophysics

## ABSTRACT

The Wide Field Infrared Survey Telescope (WFIRST) is a 2.4m space telescope with a 0.281 deg$^2$ field of view for near-IR imaging and slitless spectroscopy and a coronagraph designed for > $10^8$ starlight suppression. As background information for Astro2020 white papers, this article summarizes the current design and anticipated performance of WFIRST. While WFIRST does not have the UV imaging/spectroscopic capabilities of the Hubble Space Telescope, for wide field near-IR surveys WFIRST is hundreds of times more efficient. Some of the most ambitious multi-cycle HST Treasury programs could be executed as routine General Observer (GO) programs on WFIRST. The large area and time-domain surveys planned for the cosmology and exoplanet microlensing programs will produce extraordinarily rich data sets that enable an enormous range of Archival Research (AR) investigations. Requirements for the coronagraph are defined based on its status as a technology demonstration, but its expected performance will enable unprecedented observations of nearby giant exoplanets and circumstellar disks. WFIRST is currently in the Preliminary Design and Technology Completion phase (Phase B), on schedule for launch in 2025, with several of its critical components already in production.

## I. WFIRST

The Wide Field Infrared Survey Telescope (WFIRST) is a 2.4m telescope equipped with a large area, 300-megapixel, near-infrared camera for imaging and slitless spectroscopy and an on-axis, visible-light coronagraph for imaging, polarimetry, and low-resolution spectroscopy of circumstellar disks and nearby giant exoplanets. The 0.281 deg$^2$ field of view (FoV) is 200× that of HST's Wide Field Camera 3 near-IR channel and 90× that of HST's (optical) Advanced Camera for Surveys. The coronagraph is expected to provide several orders of magnitude greater starlight suppression than existing instruments.

WFIRST was the highest ranked large space mission in the 2010 decadal survey of astronomy and astrophysics (Astro2010; NWNH). There it was envisioned as a 1.5m telescope with a focal plane comprised of 36 H2RG (2k×2k) near-IR detectors. WFIRST was subsequently reconfigured to take advantage of a 2.4m telescope assembly made available to NASA by another federal agency, and of 18 H4RG (4k×4k) detectors, a technology that matured after Astro2010. The move to a 2.4m telescope also made WFIRST an attractive platform to advance Astro2010's highest priority medium-scale space activity -- maturing technology for direct imaging and spectroscopy of exoplanets -- by incorporating a coronagraphic instrument (CGI). This article presents a brief overview of WFIRST as background information for science white papers addressed to the Astro2020 decadal survey.

Table 1 summarizes the WFIRST instrument suite, and Figure 1 presents the effective area of the Wide Field Instrument (WFI) as a function of wavelength for the imaging filters, grism, and prism. Minor changes to these specifications are possible as the mission design evolves; in particular, the WFI prism bandpass and resolution and some CGI parameters are still being optimized. Substantial changes are not expected. Both CGI channels have a 1024 x 1024 pixel electron-multiplying CCD (EMCCD) in the focal plane. Three of the five CGI observing modes will be fully commissioned prior to launch; support for the other two modes will be on a best-effort basis. The CGI imaging modes can all be used with linear polarizers. WFIRST will operate from a Quasi-Halo orbit around the second Sun-Earth Lagrange point (L2).

## II. Wide Field Instrument (WFI): Observing Program and Sensitivity

As outlined by Astro2010, WFIRST is a survey telescope that will devote substantial fractions of its 5-year prime mission to (a) a large area, high-latitude imaging and spectroscopic survey that enables high-precision cosmological measurements with weak lensing and galaxy clustering, (b) a time-domain survey (~5 day cadence) of several tens of $deg^2$ that enables discovery and light curve monitoring of thousands of Type Ia supernovae to z ~ 2, (c) a time-domain survey (~15-minute cadence) of the Galactic bulge that enables discovery of thousands of exoplanets at AU and larger separations via gravitational microlensing, and (d) a General Observer (GO) program. Although the survey programs are designed around measurement objectives in dark energy and exoplanet demography, they will allow an enormous range of scientific investigations.

There will be no proprietary period for survey data. NASA intends to fund an Archival Researcher (AR) program to support full scientific exploitation of the WFIRST data sets. The AR program is analogous to archival research programs on HST, but it will account for a much larger fraction of WFIRST science and science funding. Appendix D of Spergel et al. (2015) describes more than 40 community-contributed ideas for WFIRST AR and GO programs, on topics that include solar system objects, exoplanet transits, brown dwarfs and stellar remnants, dynamics and resolved stellar populations of the Milky Way and nearby galaxies, galaxy evolution, quasars, gravitational lenses, and the sources of reionization. Figure 2 presents a simulated WFIRST image of a fictitious region of the sky that illustrates the extraordinary range of science opportunities afforded by its combination of HST-quality resolution with an enormous field of view.

The WFIRST formulation science investigation teams are actively investigating optimal designs and technical requirements for the surveys above, and capabilities that will maximize the science return of the AR and GO programs. Detailed survey designs and time allocations will be decided much closer to launch, incorporating broad community input. Under current plans, 25%-30% of the 5-year prime mission would be dedicated to the GO program and 5% would be devoted to coronagraphic technology demonstration. WFIRST is being designed with a 10-year mission life goal (fuel is the only expendable), and a larger fraction of time (likely 100%) of an extended mission would be operated in GO mode.

Table 2 lists characteristics of WFIRST sensitivity for a variety of representative observing modes. Yields are quoted per month of on-sky observing time, including estimated slew+settle and readout overheads but not including calibration observations, and assuming programs that are long enough to minimize "edge effect" losses in the time-domain surveys. The High Latitude Survey (HLS) is intended to provide 4-band imaging and slitless spectroscopy over a large contiguous area, with a dither pattern that visits each point in the survey 4-8 times in each band. With the currently planned exposure times (140 sec/filter/visit) it would achieve a depth (5σ point source) of 26.9 in its most sensitive bands and measure weak lensing shapes of 27 million galaxies per month ($n_{eff}$ =45 $arcmin^{-2}$). Forecasts based on empirical luminosity functions predict that the HLS will detect more than 20,000 galaxies per month at redshift $z > 8$, and about 1500 per month at $z > 10$. The spectroscopic HLS observations will reach a (5σ) emission line sensitivity of $1.0\times10^{-16}$ erg $s^{-1}$ $cm^{-2}$ at 1.8μm, near one end of the 1.0-1.93μm bandpass where the throughput is low; the forecast yield is $2.1\times10^6$ Hα redshifts ($z$ = 0.52-1.94) and $5.5\times10^5$ [OIII] redshifts ($z$ = 1.00-2.85) per month.

There are numerous parameters to consider in optimizing the supernova survey (Hounsell et al. 2018), including overlap with ground-based observations. It is likely to include at least two tiers targeting supernovae in different redshift ranges, with fields observed at a ~5 day cadence over a ~2 year span. Table 2 lists per epoch depth for one possible 2-tier configuration (areas of 18 $deg^2$ and 8 $deg^2$), and a co-added depth assuming 6 months of total observing time. A survey of this design would discover approximately 512 (956) SNIa per on-sky month in the medium (deep) tiers with photometric measurements precise enough for cosmology, a total of ~8800 SNIa to $z = 1.7$ over the full survey. These numbers assume a pure imaging strategy, but they will likely be adjusted by a trade of prism spectroscopy for some reductions in filter coverage or in survey area.

The microlensing survey plans to observe 2 $deg^2$ (7 fields) in the Galactic bulge with the 15-minute cadence needed to detect Mars-mass planetary microlensing events, over multiple seasons each lasting 72 days. The photometric precision per visit is ~ 0.001 mag at W146 ~ 16 and 0.01 mag at W146 ~ 21.2, and each field is visited ~2800 times per month. Forecasts with an up-to-date planetary population model imply detection of ~100 bound exoplanets per month of observing time (Penny et al. 2018), and about 16 free-floating Earth-mass planets per month if there is one such ejected planet per star (S. Johnson et al., in preparation). The ratio of bound to free-floating planets may be a strong discriminator of exoplanet formation models. As emphasized by Bennett et al. (2010) and Penny et al. (2018), WFIRST microlensing perfectly complements Kepler transits because it is most sensitive to planets with separations > 1 AU, including unbound planets, and to masses as low as Mars. In addition, the microlensing survey should detect about 70,000 transiting planets over six seasons, including Jupiter-sized planets with periods as long as a month, short-period planets as small as two Earth radii, and ~1200 planets with detected secondary eclipses (Montet et al. 2017). The microlensing survey will also yield superb astrometric precision, providing a unique capability for solar system, stellar, and Galactic astrophysics (Gould 2014, Gould et al. 2015).

The final two modes listed in Table 2 assume an HLS-like strategy but exposures 20× longer per visit, so that read noise is negligible and overheads are small. In this regime, a 0.5-mag increase of depth (or factor 1.6 of emission line flux sensitivity) requires a factor 2.5 increase of observing time, and a corresponding decrease in survey rate. These examples can thus be scaled to a variety of deep GO programs.

To convey the extraordinary survey power of WFIRST, it is useful to make comparisons to some of the iconic large HST programs. We include slew+settle and readout overheads for WFIRST observations where relevant and compute HST observing times (which also include overheads) by multiplying the number of orbits by 5700 sec/orbit.

*COSMOS* (Scoville et al. 2007) used 572 HST orbits (3260 ksec) to observe 1.7 $deg^2$ to a depth of 28.6 mag in I814 with ACS, producing the world's premier data set for space-based weak lensing and a major resource for AGN and galaxy evolution studies. For a typical faint galaxy color (I-H ~ 0.8) the equivalent H-band depth is 27.8 mag. Observing 1.7 $deg^2$ to this depth with WFIRST requires ≈ 26 ksec (factor of 125 speedup).

*CANDELS-Wide* (Grogin et al. 2011; Koekemoer et al. 2011) surveyed approximately 800 $arcmin^2$ to a magnitude limit J, H ≈ 27.0. The J and H imaging of CANDELS-Wide required 314 HST orbits (1790 ksec). Observing one WFIRST field (0.281 $deg^2$ ≈ 1015 $arcmin^2$) to this depth in H and J would require 1.7 ksec (factor of 1050 speedup).

*3-d HST* (Momcheva et al. 2016) used 248 orbits (1400 ksec) to observe 626 $arcmin^2$ with the G141 grism of WFC3 (and parallel G800L observations with ACS), achieving a typical 5σ line flux depth of ~ $5\times10^{-17}$ erg $s^{-1}$ $cm^{-2}$ at 1.5μm. A single field (1015 $arcmin^2$) WFIRST grism observation to this depth requires 1.9 ksec (factor of 730 speedup).

*FIGS* (Pirzkal et al. 2017) used 160 orbits (900 ksec) to survey four WFC3 fields (0.005 $deg^2$) with the grism to achieve a typical emission line sensitivity of ~ $10^{-17}$ erg $s^{-1}$ $cm^{-2}$ at $\lambda$ ~ 1μm. The WFIRST grism can reach this depth in 66 ksec at background levels typical of the High-Latitude Spectroscopic Survey at $\lambda$ ~ 1.5μm, over a broader wavelength range and higher spectral resolution, with any given 0.281 $deg^2$ pointing covering an area 55× larger than all of FIGS (assuming half of the set of roll angles is unusable for any given object as a result of contamination from nearby objects). Multiplying the increased area by the reduced exposure time implies a factor of 750 speedup.

*PHAT* (Dalcanton et al. 2012) imaged 0.5 $deg^2$ of the Andromeda galaxy in six bands; 414 orbits (2360 ksec) were used for F110W and F160W observations reaching depths of about 25.5 and 24.6, respectively, in the uncrowded outer disk. Two WFIRST pointings would cover the same area as the 414 PHAT pointings and require a total of 1.6 ksec to achieve a depth of 25.5 in each of Y, J and H (which together span the equivalent of the HST F110W and F160W filters), a factor of 1475 speedup.

As these examples make clear, routine "small" GO programs on WFIRST can easily exceed some of the most ambitious and highest impact Treasury-scale HST programs. The speedups relative to HST are often much larger than the simple factor of 200 suggested by the ratio of the WFI FoV to that of WFC3-IR, because overheads can be drastically lower, in part because of the greater efficiency of observing from L2 rather than low Earth orbit. The large WFIRST surveys will produce data sets that are hundreds or thousands of times more powerful than the largest HST-era surveys. Summaries of the current plans and anticipated performance of these surveys will appear in white papers submitted to the Astro2020 decadal survey.

While survey speed is the critical metric for many programs, we emphasize that WFIRST is also designed for the extremely stringent control of observational systematic errors needed for modern supernova and weak lensing cosmology. The optics and metering structures will be temperature controlled and will not have the disturbances present in HST's low Earth orbit, enabling a wavefront stability of < 1 nm for the WFI optical path over the course of an exposure. By incorporating recent process improvements, the WFIRST development detectors have demonstrated an order of magnitude lower persistence than the WFC3-IR detector. The onboard illumination source enables flat fields and several distinct modes to characterize detector non-linearity in flight. These requirements are driven by the dark energy programs, but a highly stable observatory will benefit a wide range of other precision applications.

## III. Coronagraphic Instrument (CGI)

The CGI is a high-contrast imager and integral field spectrograph that will enable the study of exoplanets and circumstellar disks at visible wavelengths. Ground-based high-contrast instrumentation is limited to flux ratios of ~$10^{7\text{-}8}$ at small working angles in full field of view imaging, even under optimistic assumptions for 30m-class telescopes (Guyon, 2005; Males & Guyon, 2018). Yet there is a strong scientific driver for better performance, especially at visible wavelengths. In particular, future flagship mission concepts aim to image Earth analogues with visible light flux ratios >$10^{10}$.

WFIRST CGI is a critical intermediate step toward that goal, with a predicted $10^{8\text{-}9}$ flux ratio capability. By meeting its technology demonstration requirements, CGI will enable the study of planets 20 to 100 times fainter than is possible with current ground-based coronagraphs, while simultaneously reaching more challenging shorter wavelengths. Performance *predictions* (as distinct from requirements that must be tested ahead of launch) imply an even more capable CGI, allowing the study of planets a billion times fainter than their host stars with separations as small as 0.15 arcsec. Figure 3 shows the levels of flux ratio reachable by CGI based on requirements and based on best estimates of design performance, in comparison to the capabilities of current instruments. It should be noted that CGI will only achieve its full performance on bright ($V \leq 5$) stars, and its performance on fainter stars is not yet well-constrained. Table 1 details the available bandpasses (ranging from 525nm to 825nm), coronagraphs (from 3λ/D to 20λ/D working angles), and channels (imager/polarimeter and IFS).

CGI achieves this capability through improvements over current ground and space systems in several areas:

- Hardware: space-qualified deformable mirrors, electron multiplying CCD detector (EMCCD), and two novel coronagraph types optimized for WFIRST.
- Algorithms: wavefront sensing and control at both high and low order; post-processing of high-contrast, low-flux, photon-counting EMCCD data.
- Validation of telescope and instrument models at high accuracy and precision.

As a technology demonstration, CGI will utilize these advances in space for the first time, paving the way for future missions such as HabEx or LUVOIR and dramatically reducing the implementation risk for such missions. CGI will also do groundbreaking science in its own right, in direct imaging, polarimetry, and spectroscopy of exoplanets and circumstellar disks.

CGI will receive ~3 months of observing time during the first 18 months of the mission for its "technology demonstration (TD) phase." Assuming a successful completion, NASA may make the CGI available to the general astronomy community as a science instrument. An open call for a Participating Scientist Program (PSP) is expected in ~2021. The PSP will enable members of the community to engage in the technology demonstration phase, and, if warranted by instrument performance, the PSP may perform science operations beyond the 18 month technology demonstration period.

*Exoplanets*

During the TD phase, CGI will be capable of imaging and characterization of both mature Jupiter analogues and young self-luminous giant planets. CGI is expected to image a handful of known radial velocity (RV) planets, resolving the sin(i) ambiguity and revealing their true masses, as well as distinguishing the presence or absence of cloud layers. It will also obtain a spectrum of a known Jupiter-like RV planet, covering the 730nm $CH_4$ feature, providing clues into atmospheric temperature, cloud properties, and composition when combined with 575nm broadband imaging. Performing optical spectroscopy and polarimetry on one or more self-luminous young giant planets would reveal the broad absorption features of gaseous Na and K. The abundance of these species provides additional constraints on metallicity, temperature, and gravity (and thus planet mass).

A PSP phase could address additional goals including: a blind search program for undiscovered Jupiters and possibly mini-Neptunes, comparative spectral studies of self-luminous and reflected light planets, polarimetric constraints on reflected-light Jupiters, a search for temporal variability of self-luminous planets due to patchy cloud decks, and a search for Hα emission from forming protoplanets.

While the WFIRST CGI sample will be small, it is crucial to test the end-to-end modeling and interpretation of planets detected in reflected light before much larger scale missions set out to find and characterize potentially habitable planets. Understanding the extent to which the contributions of clouds, hazes, and other effects can be discerned from reflected light data will give confidence that future missions studying more challenging targets will be properly prepared to succeed.

*Circumstellar disks*

Arguably the most important science contributions from WFIRST CGI will come from imaging and polarimetry of debris, exozodiacal, and protoplanetary disks. CGI can characterize known disks with unprecedented fidelity at small separations, in both total intensity and polarized intensity in at least two photometric bands, providing constraints on dust grain shape, composition, and grain size distribution.

Debris disks – the product of collisions between planetesimals – represent a later stage of planet formation. Many debris disks appear to be shaped into narrow rings, indicating belts of planetesimals shepherded by planets that have already formed but may be too faint for direct detection. Additionally, many recent observations of debris disks show significant temporal variability, potentially planet-induced.

CGI will be capable, for the first time, of imaging tenuous habitable-zone interplanetary dust (exozodiacal dust). For nearby stars, CGI will be sensitive to dust densities only ~10x greater than those in the Solar System, reaching the regime where disk structure is dominated by transport phenomena rather than collisions.

Imaging of protoplanetary disks and protoplanets yields important constraints on when and where planets form. Many protoplanetary disks feature large scale structures, some of which may be explained by the formation and growth of planets. Detection of localized Hα emission could help distinguish protoplanets from other clumpy disk features. A challenge will be the CGI constraint on host star magnitude for best starlight rejection performance ($V \leq 5$), with the performance on fainter stars yet to be determined.

During the TD phase, CGI will conduct dual-band imaging and polarimetry of several known debris disks. In this phase, it may also conduct a search for exozodi around a high priority target. PSP campaigns could encompass imaging and polarimetry of protoplanetary disks, monitoring for temporal variability of disk structures, comparative studies of debris and protoplanetary disks, and vetting of prime targets for future exo-Earth imaging missions to ensure they are not contaminated by bright exozodiacal disks.

## IV. WFIRST Status

WFIRST is currently in Phase B, Preliminary Design and Technology Completion, working towards a preliminary design review and entry to Phase C in Fall of 2019. NASA has imposed a cost cap of $3.2B on the Phase B mission design (full life-cycle cost, as-spent dollars). The increase relative to the $1.6B (FY2010 dollars) quoted by Astro2010 can be divided approximately among inflation ($0.7B), the addition of the coronagraph ($0.5B), the addition of GO program funding ($0.1B), and $0.3B in cost growth, associated partly with the change to the 2.4m mirror. Meeting the $3.2B cost cap has required a number of descopes relative to the mission concept presented in Spergel et al. (2015), most significantly the removal of the WFI integral field channel, removal of some coronagraph operation modes, and change of the coronagraph status from science instrument to technology demonstration with associated acceptance of increased risk.

Key contracts were signed in 2018 for the WFI optical/mechanical assembly and the telescope. Production of the flight infrared detectors is in progress, with the first flight candidate detectors scheduled to undergo acceptance testing in spring 2019. Re-figuring of the primary mirror for its new application began in fall 2018. Procurements have begun for components of engineering test units, and in some cases for spacecraft flight hardware where common-buys with other missions would lead to cost savings.

The project is on schedule for a launch in 2025, with 9 months of built-in margin. Schedule and cost projections assume an optimal funding profile, and uncertainty in the funding profile is the principal source of uncertainty in the mission schedule and total cost. The Formulation Science Teams have a 5-year term, and NASA anticipates a new competition for Implementation Science Teams and Participating Scientist Teams following the critical design review partway through Phase C.

## V. 100 Hubbles for the 2020s

WFIRST combines essential strengths of the Hubble Space Telescope and the Sloan Digital Sky Survey (SDSS), arguably the two most influential astronomical facilities of the past half-century. WFIRST does not have all the capabilities of HST, most notably in UV imaging and UV/visible spectroscopy. However, as illustrated in §II, for many observing programs WFIRST is literally hundreds of times more powerful than HST, so the seemingly hyperbolic title of this section in fact understates WFIRST's capabilities. WFIRST is an ideal complement to JWST, capable of discovering rare systems that can be characterized comprehensively with JWST's much greater sensitivity and spectroscopic capability. The WFIRST coronagraph will pioneer the technology for future missions to image Earth-like worlds, and it will enable unprecedented high-resolution views of giant exoplanets and proto-planetary disks. Like other ambitious survey facilities such as the SDSS and the Large Synoptic Survey Telescope, WFIRST is designed with specific science objectives in mind, but the ultimate return from the mission will include numerous investigations and discoveries that we cannot even conceive today, touching all areas of astrophysics from the solar system to the most distant reaches of the universe.

We thank the many, many individuals who have contributed to the WFIRST mission and the WFIRST Science Definition Teams and Science Formulation Teams over the past decade. Their input has been essential to designing the mission summarized here. We also thank those in the astronomical community, NASA, Congress, and the Executive Branch whose support has allowed WFIRST to reach its current stage as a mission completing design and beginning construction for a mid-2020s launch.

## References Cited

National Research Council 2010, *New Worlds, New Horizons in Astronomy and Astrophysics,* The National Academies Press, Washington DC (NWNH)

Bennett, D. et al. 2010, *Completing the Census of Exoplanets with the Microlensing Planet Finder (MPF)*, RFI response for Astro2010, arXiv:1012.4486

Dalcanton, J. et al. 2012, *The Panchromatic Hubble Andromeda Treasury*, ApJS 200, 18, arXiv:1204.0010

Gould, A. 2014, *WFIRST Ultra-Precise Astrometry I: Kuiper Belt Objects*, JKAS 47, 279, arXiv:1403.4241

Gould, A., Huber, D., Penny, M., & Stello, D. 2015, *WFIRST Ultra-Precise Astrometry II: Asteroseismology*, JKAS 48, 93, arXiv:1410.7395

Grogin, N. et al. 2011, *CANDELS: The Cosmic Assembly Near-infrared Deep Extragalactic Legacy Survey*, ApJS 197, 35, arXiv:1105.3753

Guyon, O. 2005, *Limits of Adaptive Optics for High-Contrast Imaging,* ApJ, 629, 592, arXiv:astro-ph/0505086

Hounsell, R. et al. 2018, *Simulations of the WFIRST Supernova Survey and Forecasts of Cosmological Constraints,* ApJ 867, 23, arXiv:1702.01747

Koekemoer, A. M. et al. 2011, *CANDELS: The Cosmic Assembly Near-infrared Deep Extragalactic Legacy Survey: The Hubble Space Telescope Observations, Imaging Data Products, and Mosaics,* ApJS 197, 36, arXiv:1105.3754

Males, J. R. & Guyon, O. 2018, *Ground-based adaptive optics coronagraphic performance under closed-loop predictive control,* Journal of Astronomical Telescopes, Instruments, and Systems, 4, 019001, arXiv:1712.07189

Momcheva, I. et al. 2016, *The 3D-HST Survey: Hubble Space Telescope WFC3/G141 Grism Spectra, Redshifts, and Emission Line Measurements for ~ 100,000 Galaxies*, ApJS 225, 27, arXiv:1510.02106

Montet, B. T., Yee, J. C., & Penny, M. T. 2017, *Measuring the Galactic Distribution of Transiting Planets with WFIRST,* PASP, 129, 04401, arXiv:1610.03067

Penny, M. et al. 2018, *Predictions of the WFIRST Microlensing Survey I: Bound Planet Detection Rates,* arXiv:1808.02490

Pirzkal, N. et al. 2017, *FIGS—Faint Infrared Grism Survey: Description and Data Reduction*, ApJ 846, 84, arXiv:1706.02669

Scoville, N. et al. 2007, *The Cosmic Evolution Survey (COSMOS): Overview*, ApJS 172, 1, arXiv:astro-ph/0612305

Spergel, D. et al. 2015, *Wide-Field InfrarRed Survey Telescope-Astrophysics Focused Telescope Assets WFIRST-AFTA 2015 Report*, arXiv:1503.03757

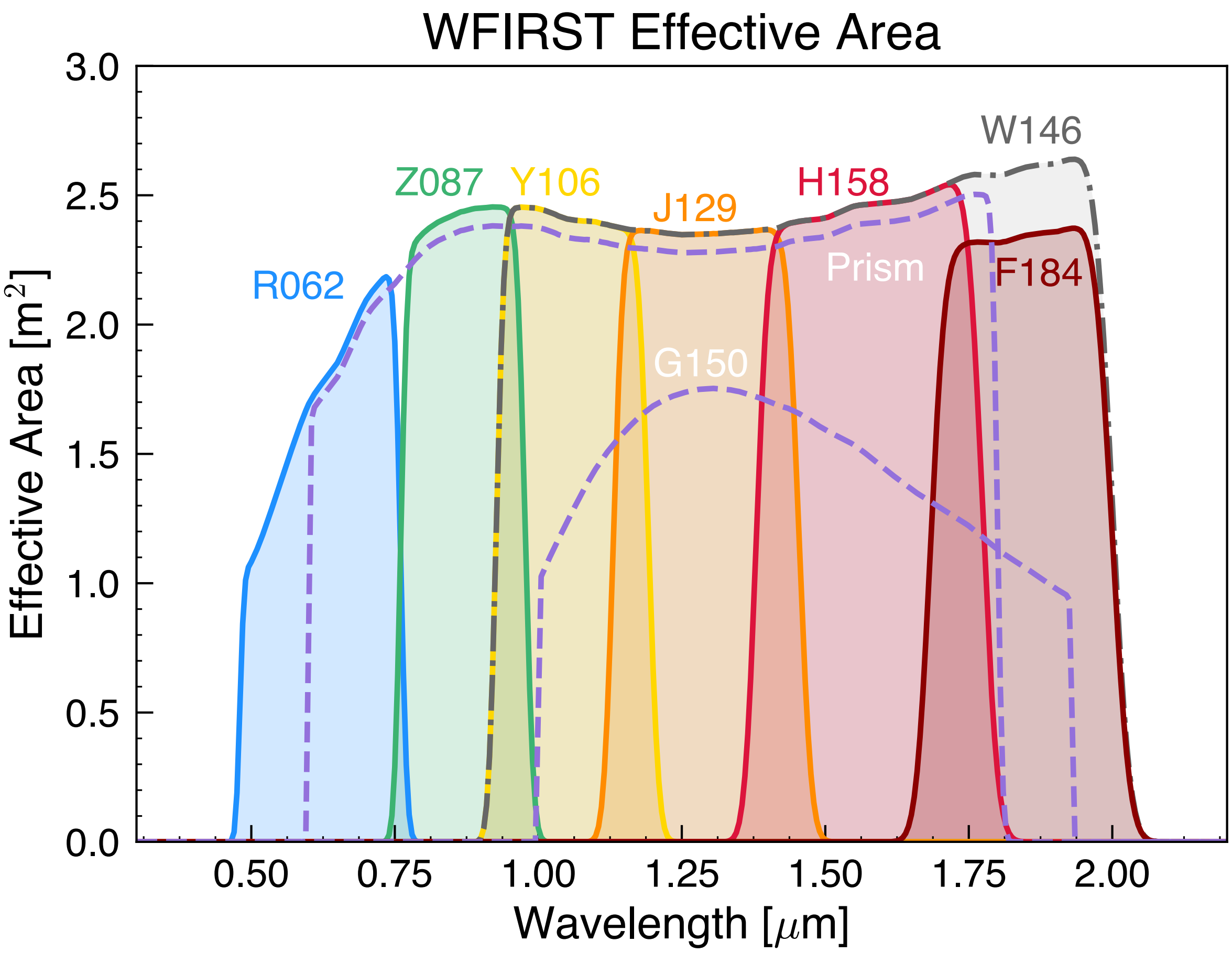

*Figure 1: Effective area of the WFI in the seven imaging filters, the grism, and the prism. Note that the W146 filter used for the microlensing survey extends from the blue edge of the Y106 filter to the red edge of the F184 filter. For comparison, the HST ACS gives an effective area of ~1.9 m$^2$ at optical wavelengths and HST WFC3/IR gives an effective area of ~2.4 m$^2$ at near-IR wavelengths.*

*Figure 2 (NEXT PAGE): A fictitious image that illustrates WFIRST's unique combination of wide area and exquisite resolution. Red squares outline the footprints of the 18 H4RG detectors that make up the 0.3-Gigapixel near-infrared camera, with a total field of view of 0.281deg$^2$; the largest zoom panel is 1.5 arc-minutes on a side and the others are 1 arc-minute. Image simulations, based on HST data, assume 1-hour exposures in the H158, Y106, and Z087 bands (red, green, blue), approximately matching the depth of the HST CANDELS survey, which forms the image background. The area of a single WFIRST pointing is 25% larger than that of the entire CANDELS-Wide Multi-Cycle Treasury Program. The large galaxy is M83; one-hour exposures readily yield resolved stellar populations at its distance of 4.5 Mpc. The globular cluster is M54 at a distance of 27 kpc; WFIRST imaging resolves bright stars even in the dense cluster core. The lower right zoom shows a hypothetical dwarf galaxy, modeled on Draco, at a distance of 3 Mpc. The top right zoom shows a region of the CANDELS-based background. For comparison, yellow and orange squares show the field of view of HST's optical Advanced Camera for Surveys and near-IR Wide Field Camera 3. Image Credit: B. Williams, E. Bell, R. Khan, D. Sand, R. Sanderson, A. Seth, D. Benford.*

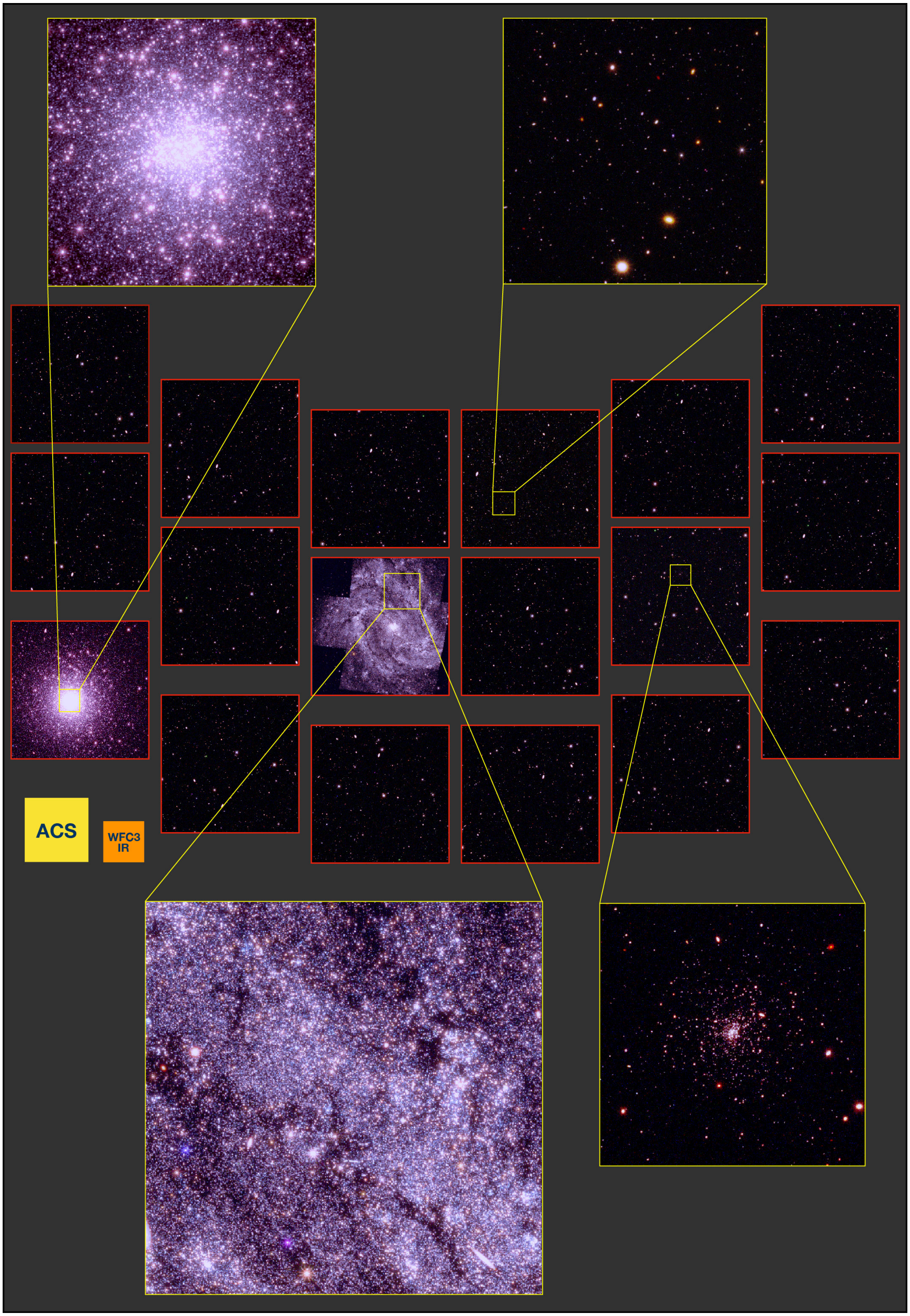
ACS
WFC3 IR

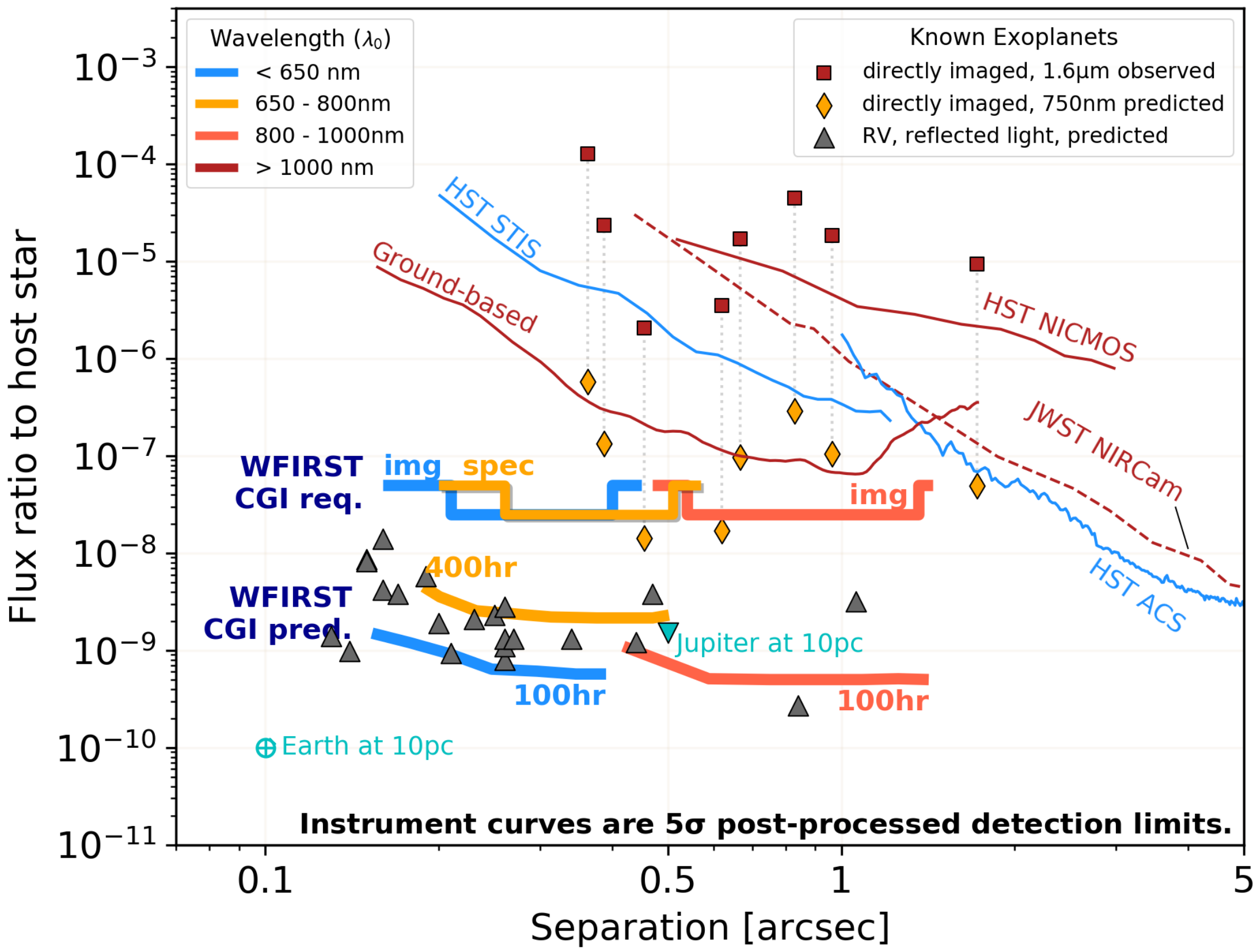

*Figure 3: Anticipated flux ratio performance of the WFIRST CGI based on the technology demonstration requirements (upper lines) and predictions for a V=5 star with the current instrument design (lower lines). Performance of existing ground- and space-based coronagraphs and forecast performance of the JWST NIRCam coronagraph are shown with lighter curves for comparison. The technology demonstration requirements are 1-2 orders-of-magnitude more sensitive than current capabilities, sufficient to detect known self-luminous, directly imaged planets at shorter wavelengths. The predicted performance is a further 1-2 orders-of-magnitude more sensitive, sufficient to detect many known RV planets in reflected light. Image Credit: V. Bailey.*

**Table 1: WFIRST Instrumentation**

| Wide-Field Instrument | | | | | | | |
|---|---|---|---|---|---|---|---|
| Focal plane: 18 $4088^2$ HgCdTe H4RG | | | | Pixel scale: 0.11” | | FoV: 0.281 sq deg | |
| Point-source Imaging sensitivity, 5σ, in 1 hour | | | | | | | |
| Filter | R062 | Z087 | Y106 | J129 | H158 | F184 | W146 |
| Bandpass (μm) | 0.48-0.76 | 0.76-0.98 | 0.93-1.19 | 1.13-1.45 | 1.38-1.77 | 1.68-2.0 | 0.93-2.0 |
| AB mag | 28.5 | 28.2 | 28.1 | 28.0 | 28.0 | 27.5 | 28.3 |

| Point-source slitless spectroscopic sensitivity, continuum mag (AB)10σ/pix, in 1 hr | | | | | |
|---|---|---|---|---|---|
| | Bandpass μm | Resolution | 0.8 μm | 1.1 μm | 1.5 μm |
| Grism | 1.0-1.93 | 435-865 | N/A | 20.78 | 20.48 |
| Prism | 0.6-1.8 | 70-140 | 22.4 | 22.75 | 22.65 |

| Coronagraph Instrument | | | | | | | |
|---|---|---|---|---|---|---|---|
| Direct Imaging Camera | | | Pixel scale: 0.021” | | | FoV: 9” x 9” | |
| Integral Field Spectrograph | | | Lenslet spacing: 0.029” | | | FoV: 2.2” x 2.2” | |
| Coronagraph Observing modes | | | | | | | |
| Filter | $\lambda_{center}$ (nm) | Band-width | Mask Type | Channel | Sup-ported ? | Working Angle | Starlight Suppression Region |
| 1 | 575 | 10% | HLC | Imager | Y | 3-9 λ/D | 360° |
| 2 | 660 | 15% | SPC “bowtie” | IFS | | 3-9 λ/D | 130° |
| 3 | 730 | 15% | SPC “bowtie” | IFS | Y | 3-9 λ/D | 130° |
| 4 | 825 | 10% | SPC wide FOV | Imager | Y | 6.5-20 λ/D | 360° |
| 4 | 825 | 10% | HLC | Imager | | 3-9 λ/D | 360° |

**Table 2: WFI Observing Sensitivity**

| *Observing Mode* | *Sensitivity* | *Yield per On-Sky Month* |
|---|---|---|
| *HLS Imaging*<br>Wide-area, 4-band | Y=26.9 J=26.95<br>H=26.9, F184=26.25 | 170 $deg^2$<br>$2.7\times10^7$ WL shapes |
| *HLS Spectroscopy*<br>Slitless, $\lambda$ = 1.0-1.93μm | $1.0\times10^{-16}$ erg $s^{-1}$ $cm^{-2}$<br>@ 1.80 μm | 303 $deg^2$<br>$2.6\times10^6$ redshifts |
| *Supernova Deep Imaging*<br>5-day cadence, 8.5 $deg^2$<br>Z087, Y106, J129, H158, F184 | 1 epoch: Z=25.9, Y=26.7, J=26.6, H=26.8, F=27.0<br>6-month coadd:<br>Z=28.7,Y=29.5, J=29.4, H=29.6, F=29.7 | 956 SNIa |
| *Supernova Wide Imaging*<br>5-day cadence, 18.5 $deg^2$<br>R062, Z087, Y106 ,J129, H158 | 1 epoch: R=25.3, Z=25.0, Y=25.0, J=25.5, H=26.0<br>6-month coadd: R=28.1, Z=27.8, Y=27.8, J=28.3, H=28.7 | 512 SNIa |
| *Microlensing*<br>2 $deg^2$, imaging<br>W146 @ 15-min cadence<br>Z087 @ 12-hour cadence | single-epoch precision:<br>0.001 mag @ W146 ~ 16<br>0.01 mag @ W146 ~ 21 | 2800 exposure/field<br>85 planets $M>1M_E$<br>15 planets $M<1M_E$ |
| *Deep Single-Band Imaging*<br>Contiguous, dithered | Any one of:<br>R062=29.6, Z087=29.23, Y106=29.15, J129=29.07, H158=29.1, F184=28.51 | 40 $deg^2$ |
| *Deep Grism Spectroscopy*<br>Contiguous, dithered | $1.0\times10^{-17}$ erg $s^{-1}$ $cm^{-2}$<br>@ 1.5 μm | 12 $deg^2$ |

***Notes:*** Yields are per month of on-sky time, including slew+settle overheads but not calibration overheads. Photometric limits are 5σ point source AB magnitudes for imaging, emission-line flux for spectroscopy.